\documentclass[twocolumn,showpacs,prl,aps,amssymb,superscriptaddress]{revtex4}

\usepackage{graphicx}
\usepackage{bm}
\usepackage{color}

\begin{document}

\title{
Connecting distant ends of one-dimensional critical systems 
by a sine-square deformation
}
\author{Toshiya Hikihara}
\affiliation{Department of Physics, Hokkaido University,
Sapporo 060-0810, Japan}
\author{Tomotoshi Nishino}
\affiliation{Department of Physics, Graduate School of Science, 
Kobe University, Kobe, 657-8501, Japan}

\date{\today}

\begin{abstract}
We study the one-dimensional quantum critical spin systems with the sine-square deformation, in which the energy scale in the Hamiltonian at the position $x$ is modified by the function $f_x = \sin^2\left[\frac{\pi}{L}(x-\frac{1}{2})\right]$, where $L$ is the length of the system.
By investigating the entanglement entropy, spin correlation functions, and wave-function overlap, we show that the sine-square deformation changes the topology of the geometrical connection of the ground state drastically; 
Although the system apparently has open edges, the sine-square deformation links those ends and realizes the periodic ground state at the level of wave function.
Our results propose a new method to control the topology of quantum states by energy-scale deformation.

\end{abstract}


\maketitle


{\it Introduction: }
Topology is one of the most fundamental concept in physics.
It rules the connectivity of local elements of the system and governs how physical objects -- particles, excitations, and information -- propagate.
Normally, the topology of a system is fixed once the spatial geometry of elements is given.
Search for other paths to the control of topology of the system is a challenging problem.

In a finite system, a boundary condition determines the topology of the geometrical connection of quantum state and affects crucially the properties of the system.
If the system has open edges, they usually induce boundary oscillations such as Friedel oscillation.
While the boundary oscillation contains important information such as the Fermi momentum, it is often regarded as an obstacle to mask the bulk properties.
One simple way to remove it is to employ the periodic boundary condition, however, there has also been another attempts, called the smooth boundary condition, to suppress the boundary effects by turning off the energy scale of local Hamiltonians smoothly around the open edges.\cite{VekicW1993,VekicW1996}
The latter has proven to be useful when the open system is favored, {\it e.g.}, for an efficiency of numerical methods such as the density-matrix renormalization group (DMRG) method.

\begin{figure}
\begin{center}
\includegraphics[width=60mm]{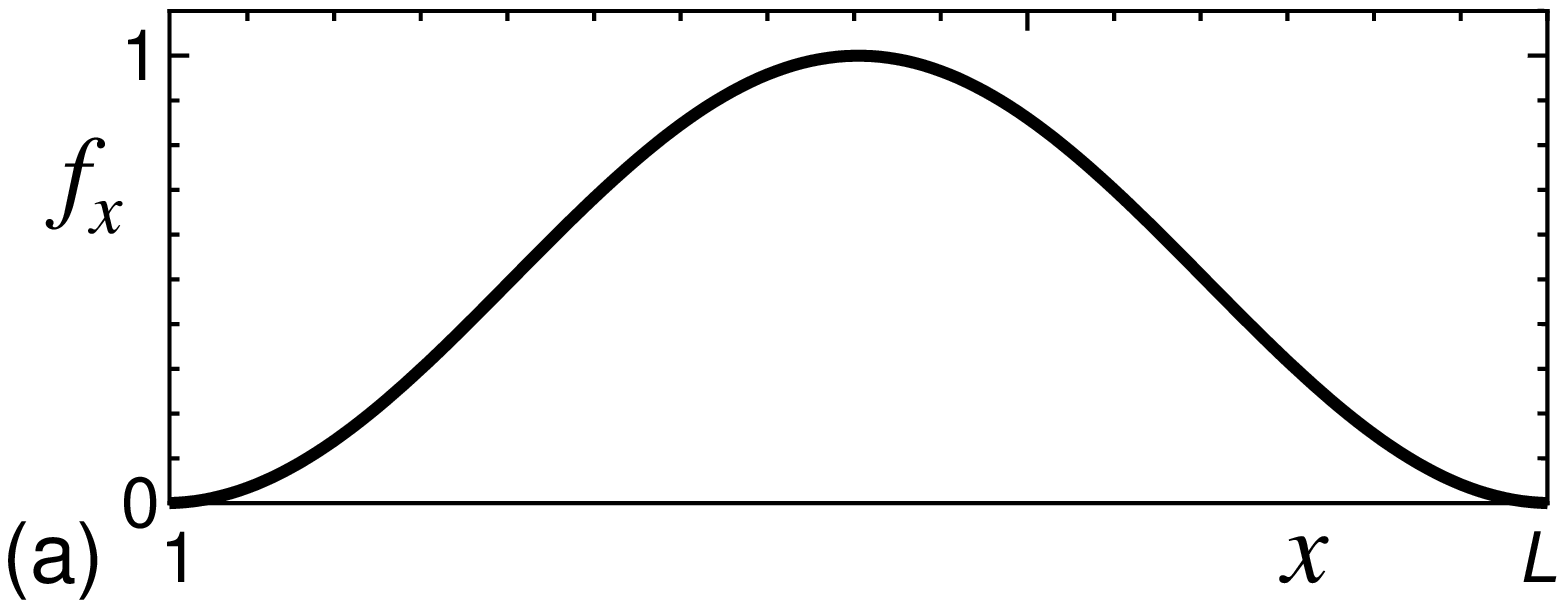}
\includegraphics[width=60mm]{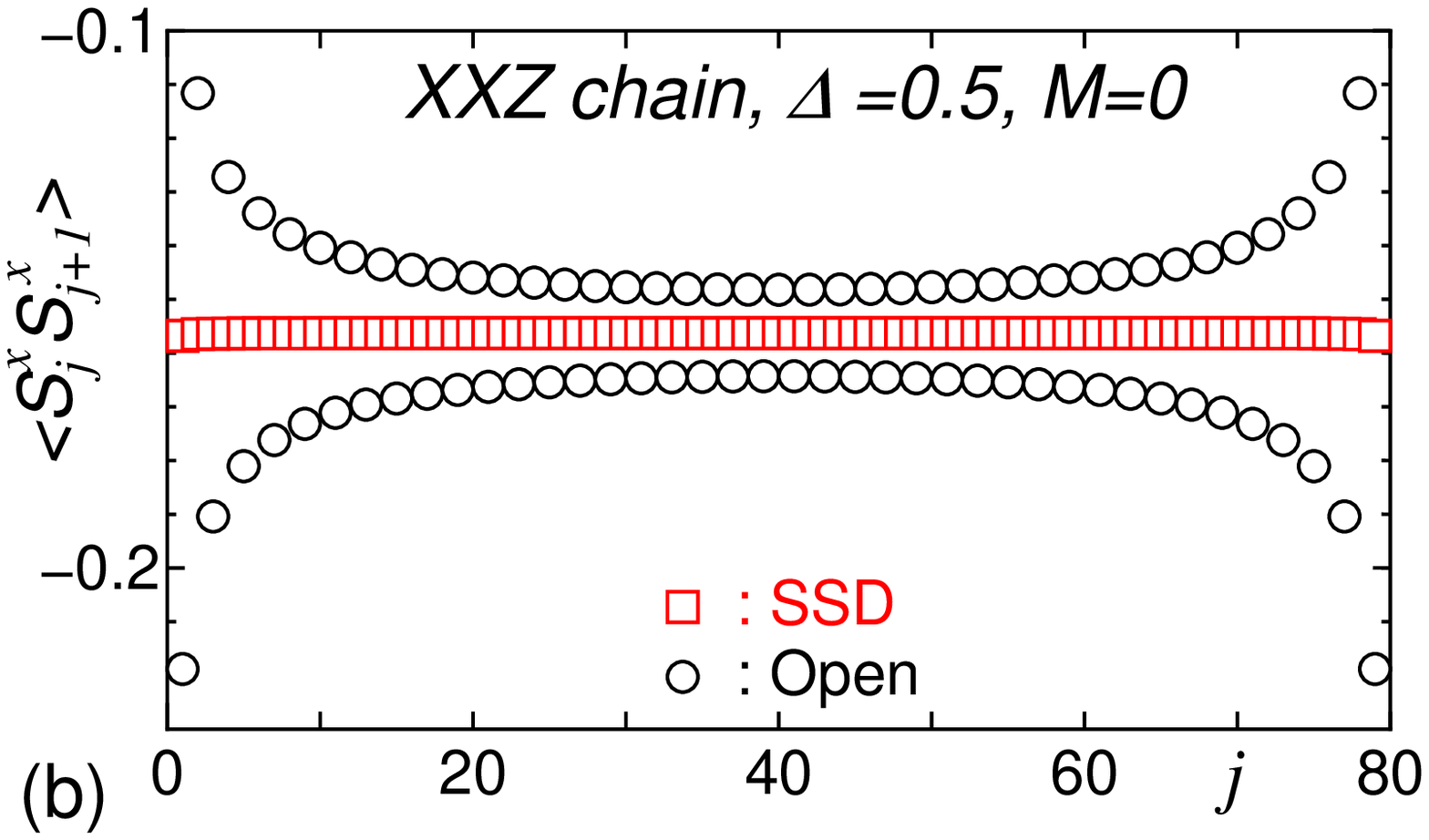}
\caption{
(a) Rescaling function $f_x$ of the SSD.
(b) Bond strength $\langle S^x_j S^x_{j+1} \rangle$ for the $XXZ$ chain (\ref{eq:Ham-XXZ}) with $L=80$ and $(\Delta, M)=(0.5, 0)$. 
Squares and circles represent the data for the chain with SSD and the uniform open chain, respectively.
}
\label{fig:Sine}
\end{center}
\end{figure}

Recently, a new scheme of the smooth boundary condition, which we call the sine-square deformation (SSD), has been proposed as an efficient way to suppress the finite-size and open-boundary effects.\cite{GendiarKN2009}
In the system with SSD, the energy scale in the Hamiltonians are modified according to the function,
\begin{eqnarray}
f_x = \sin^2\left[ \frac{\pi}{L} \left( x - \frac{1}{2} \right) \right],
\label{eq:fx}
\end{eqnarray}
where $x$ is the position of the local term and $L$ is the length of the system.
In Ref.\ \onlinecite{GendiarKN2009}, Gendiar {\it et al.} applied the SSD to the one-dimensional (1D) free fermion system with open boundaries.
They then showed that the SSD removed boundary effects successfully and resulted in position-independent one-point functions such as the bond strength and particle density in the ground state.
Since the spatial profiles of these quantities were {\it nearly completely} flat, the observation raised a natural question of what happened in the ground state of the system with SSD.
This is indeed the motivation of the present study.

In this paper, we study the SSD in several 1D quantum spin systems.
Using the DMRG and exact diagonalization methods, we study numerically the entanglement entropy (EE), correlation functions, and wave-function overlap in the systems with SSD. 
We then show that the ground state of a critical system with SSD is equivalent to that of the uniform periodic system; 
the SSD changes the topology of the critical ground state drastically, from an open chain to a periodic ring.
The result opens the possibility to control the topology of quantum states by the energy-scale deformation even in the case that the geometrical shape of the system is fixed.


{\it Sine-square deformation: }
The SSD introduces a spatial modulation of energy scale by applying the rescaling factor $f_x$ [Eq.(\ref{eq:fx})] to the local Hamiltonian at the position $x$.
For example, the model Hamiltonian of the spin-1/2 antiferromagnetic $XXZ$ chain with SSD is given by 
\begin{eqnarray}
\mathcal{H}_{\rm XXZ} &=& J \sum_{j=1}^{L-1} 
f_{j+\frac{1}{2}}
\left( S^x_j S^x_{j+1} + S^y_j S^y_{j+1} + \Delta S^z_j S^z_{j+1} \right)
\nonumber \\
&&- h \sum_{j=1}^L f_j S^z_j,
\label{eq:Ham-XXZ}
\end{eqnarray}
where we have introduced the magnetic field $h$ which induces magnetization $M$ per spin\cite{M-hrelation}.
Hereafter, we consider the case of even $L$ otherwise mentioned.
The energy scale of the local Hamiltonians thus decreases smoothly as being closer to the boundaries and eventually vanishes at the open ends, as shown in Fig.\ \ref{fig:Sine} (a).\cite{PBC-SSD}

Figure\ \ref{fig:Sine} (b) gives the DMRG data of the bond strength $\langle S^x_j S^x_{j+1}\rangle$ in the ground state of the $XXZ$ chain (\ref{eq:Ham-XXZ}) 
with and without SSD.
The data clearly show that the SSD eliminates the Friedel oscillation seen in the uniform open chain almost completely\cite{GendiarKN2009}.
We will demonstrate below that the SSD is not only an efficient measure to suppress the boundary effects but also a device to drastically change the topology of the ground-state wave function.

{\it Entanglement entropy: }
We first investigate EE in the ground state of the 1D systems with SSD. 
We consider EE for a subsystem $\Omega$ of the left $l$ spins, 
\begin{eqnarray}
\mathcal{S}(l) = - {\rm Tr}_{\Omega} \left[\rho(l) \ln \rho(l) \right],
\label{eq:EEnt}
\end{eqnarray}
where $\rho(l)$ is the reduced density matrix for $\Omega$.
For the 1D critical uniform systems, EE is known to take a universal form,\cite{Holzhey1994,Vidal2003,Calabrese2004}
\begin{eqnarray}
\mathcal{S}(l) = s \ln[g(l)] + {\rm const.},
\label{eq:CarabC}
\end{eqnarray}
where $g(l) = \frac{L}{\pi} \sin\left(\frac{\pi l}{L}\right)$.
The slope $s$ is determined by the boundary condition; $s = c/3$ for the periodic system while $s = c/6$ for the system with open boundaries, where $c$ is the central charge.
Namely, the slope $s$ divided by $c/6$ gives the number of ``cuts'' of the 1D critical state between the subsystem $\Omega$ and the environment $\bar{\Omega}$.

\begin{figure}
\begin{center}
\includegraphics[width=40mm]{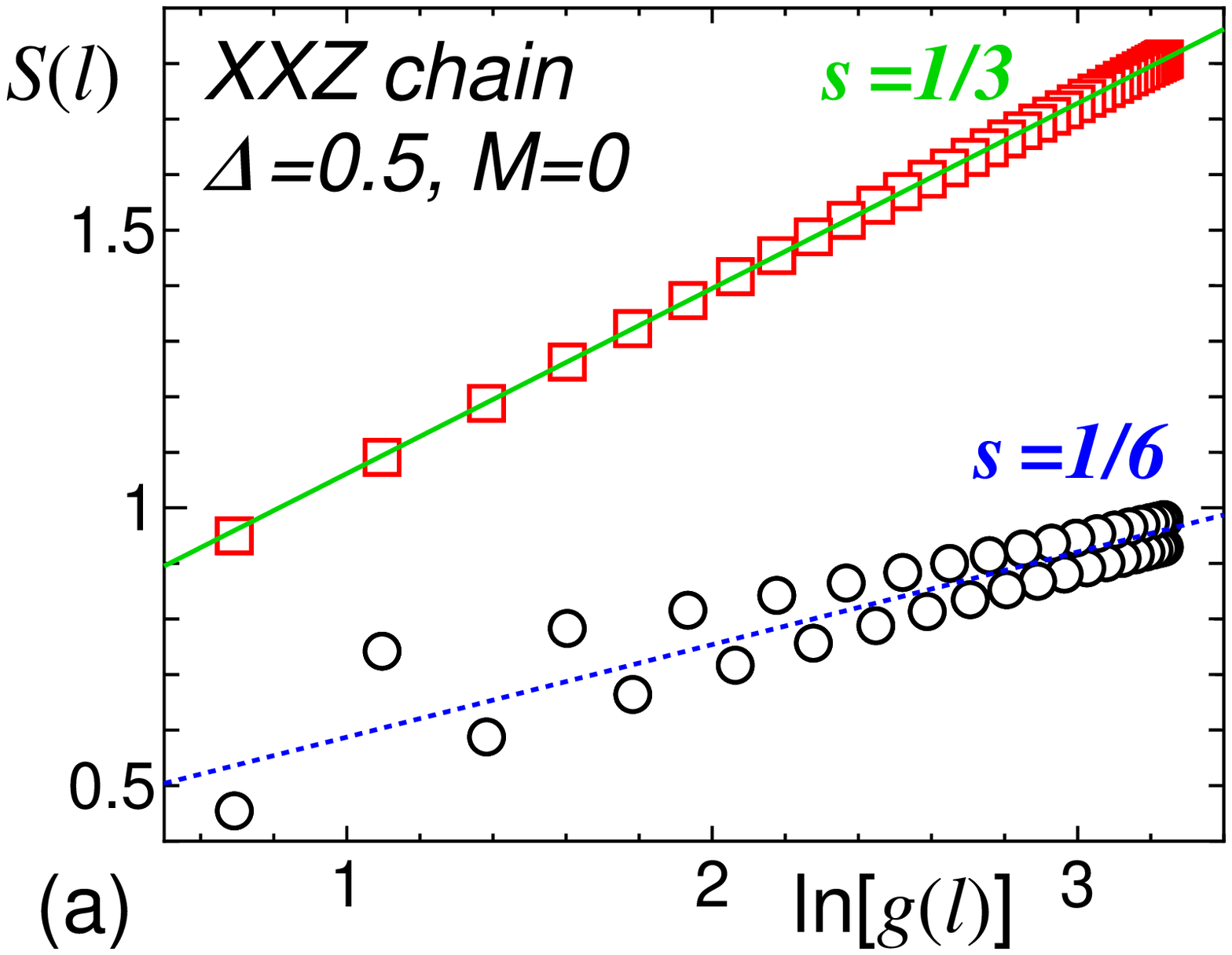}
\includegraphics[width=40mm]{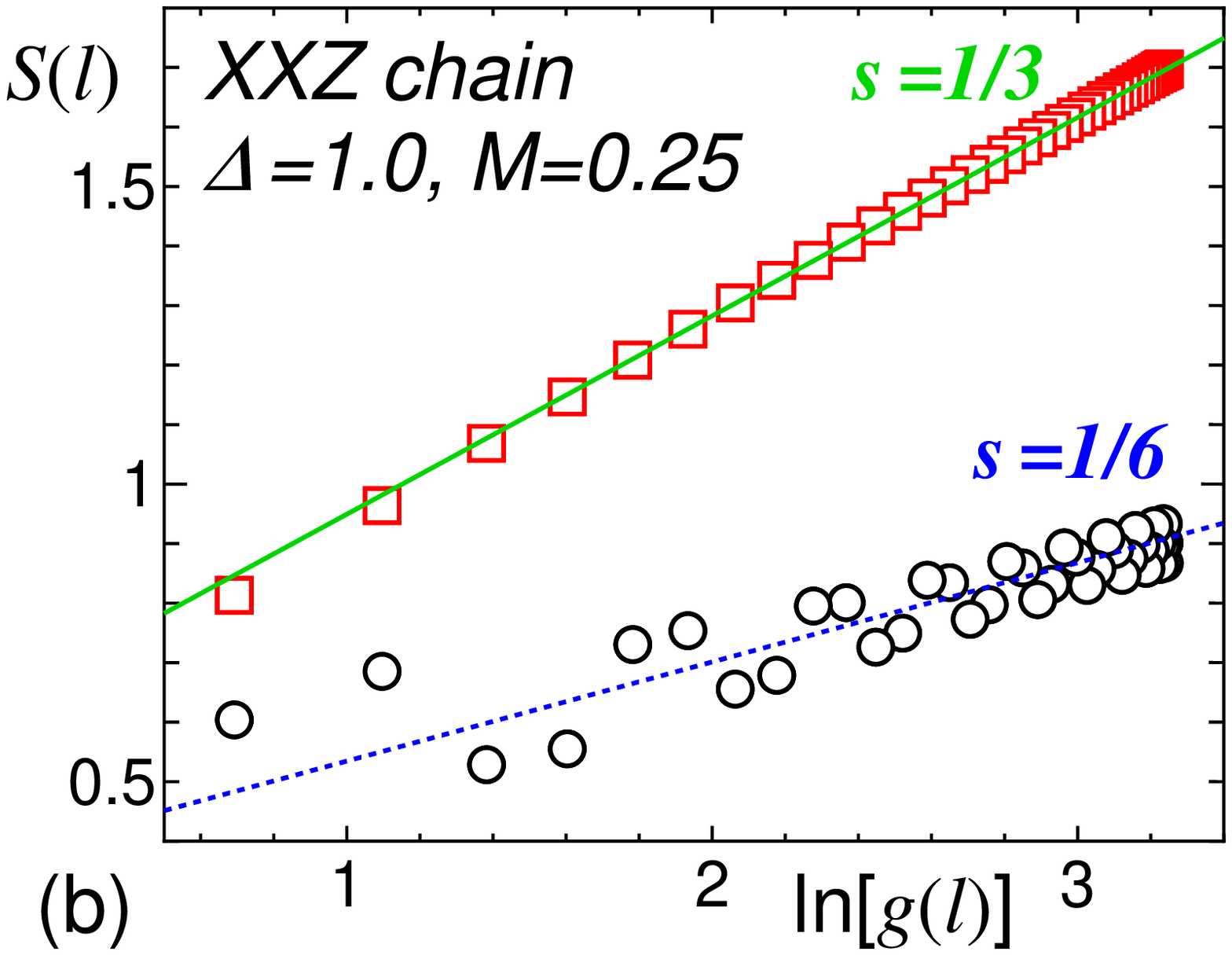}
\includegraphics[width=40mm]{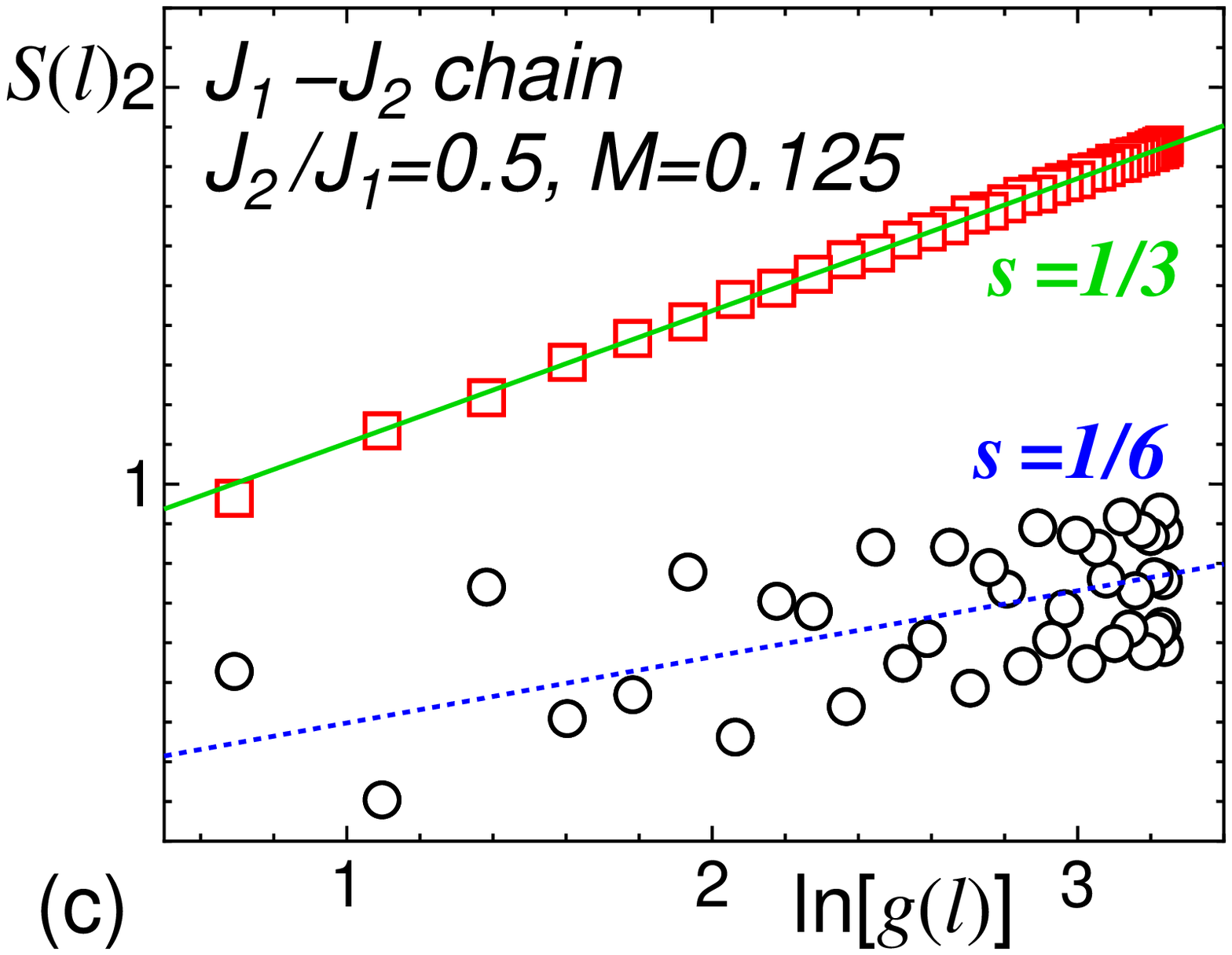}
\includegraphics[width=40mm]{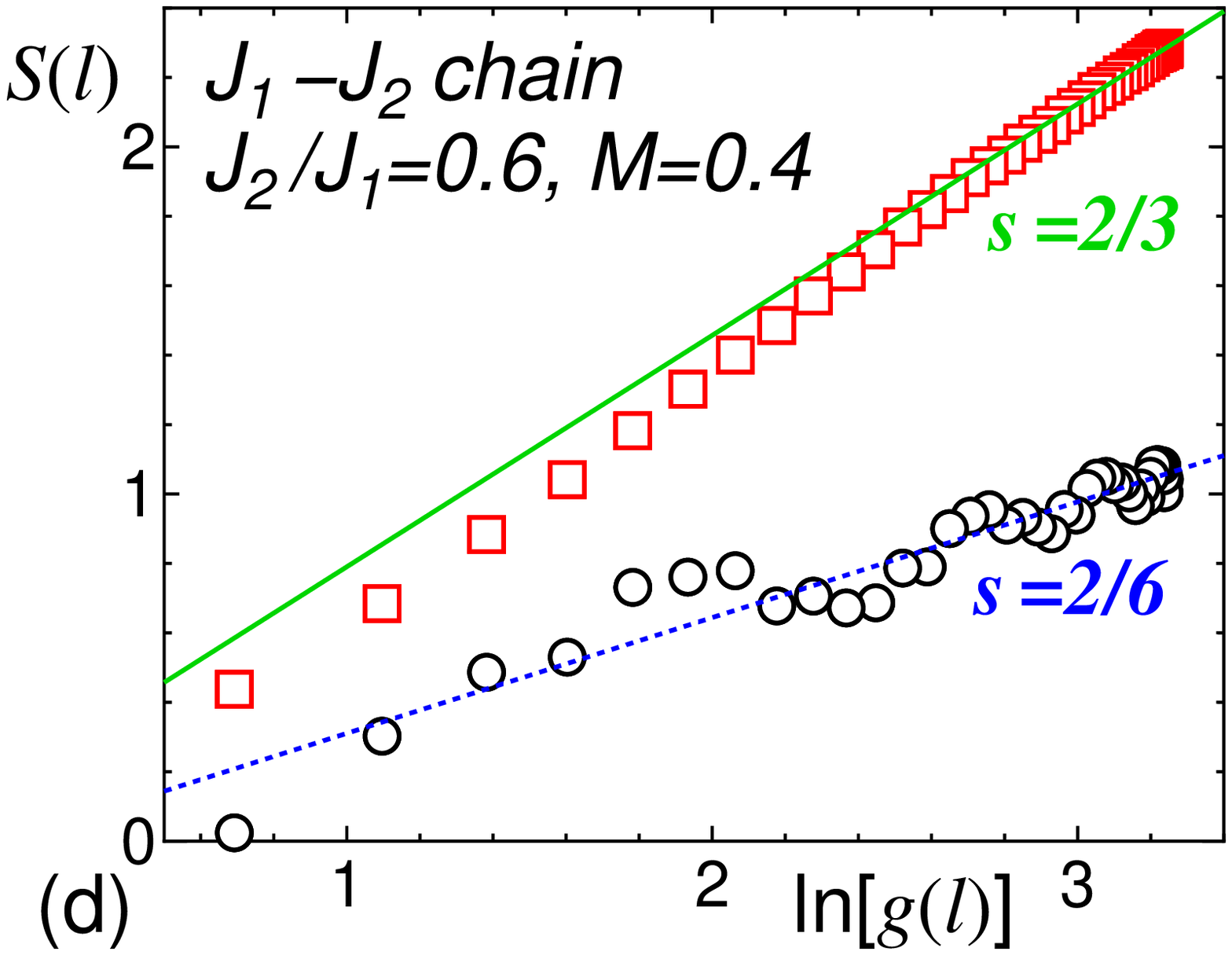}
\includegraphics[width=40mm]{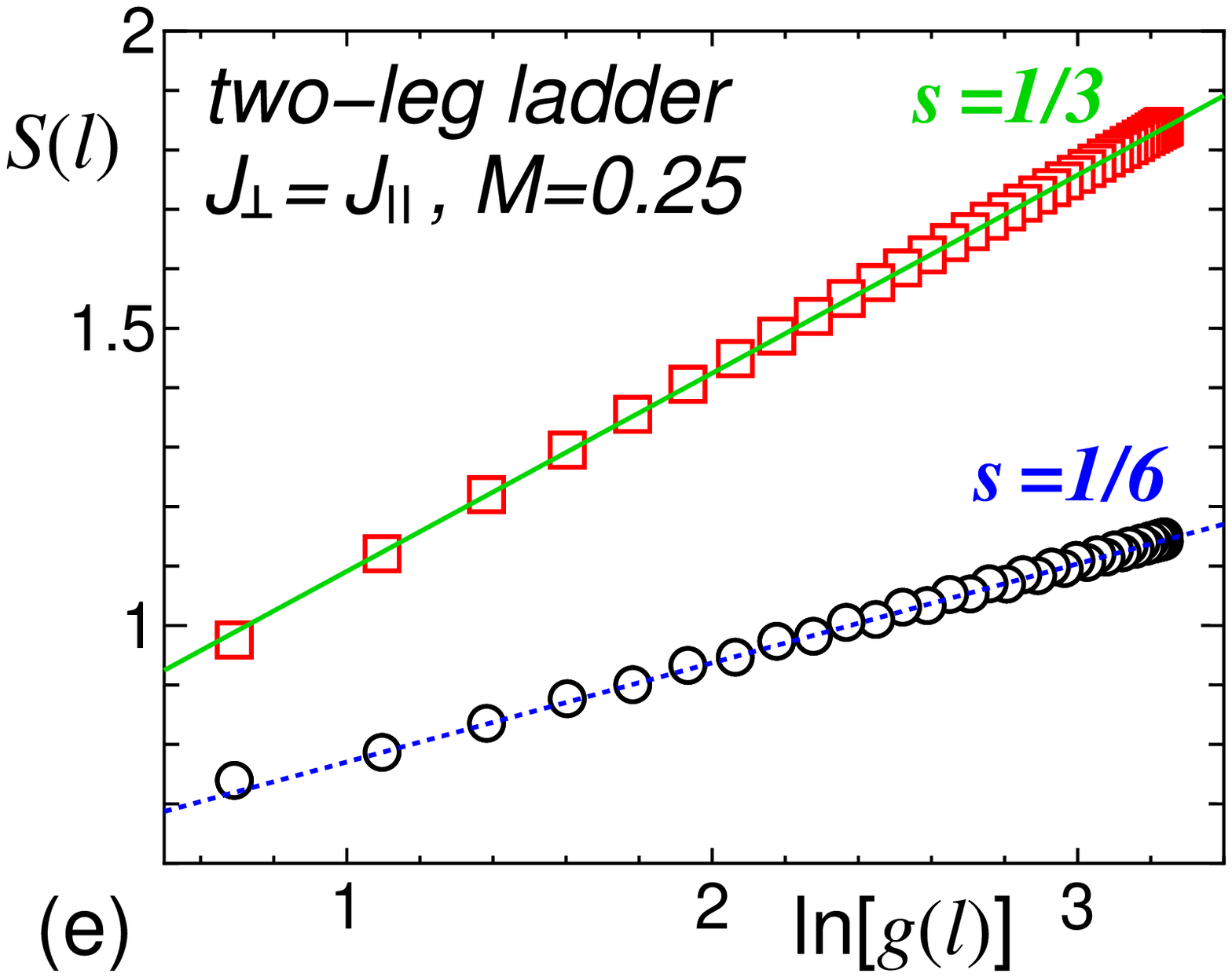}
\includegraphics[width=37mm]{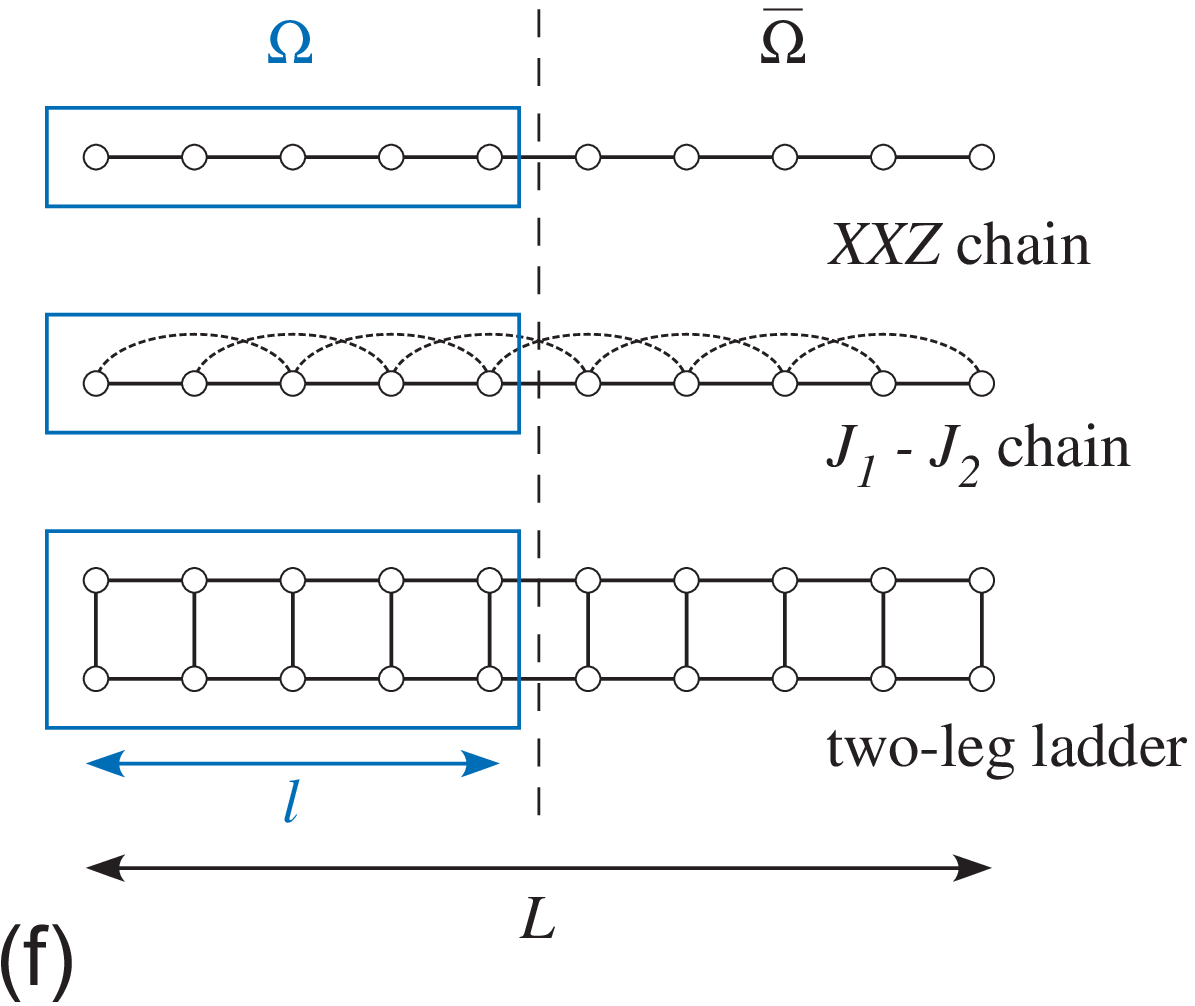}
\caption{
Entanglement entropy $\mathcal{S}(l)$ as a function of $g(l) = (L/\pi) \sin(\pi l/L)$ for $L=80$ and 
(a) $XXZ$ chain with $(\Delta, M)=(0.5, 0)$, 
(b) $XXZ$ chain with $(\Delta, M)=(1.0, 0.25)$, 
(c) $J_1$-$J_2$ chain with $(J_2/J_1, M)=(0.5, 0.125)$,
(d) $J_1$-$J_2$ chain with $(J_2/J_1, M)=(0.6, 0.4)$,
and (e) two-leg ladder with $(J_\perp/J_\parallel, M)=(1.0, 0.25)$.
The central charge $c=1$ for the models in (a)-(c) and (e), while $c=2$ for (d).
Squares and circles represent the data for the open system with SSD and the uniform open system, respectively.
Solid and dotted lines show the slope of $c/3$ and $c/6$, respectively.
(f) Shape of the subsystem $\Omega$ for which $\mathcal{S}(l)$ is calculated.
}
\label{fig:EEnt}
\end{center}
\end{figure}

Figures\ \ref{fig:EEnt} (a) and (b) show the DMRG data of EE in the $XXZ$ chain with SSD [Eq.\ (\ref{eq:Ham-XXZ})].
EE in the uniform open chains is also shown for a comparison.
Remarkably, EE in systems with SSD has the slope $s=c/3$, which is twice as large as that in the uniform open systems.
This means that the ground state of the system with SSD has two cuts between the left and right subsystems, $\Omega$ and $\bar{\Omega}$, although the lattice has seemingly only one cut.
In addition, the boundary oscillation, which is pronounced in the uniform systems, is removed by the SSD.
The results suggest that, although the system apparently possesses the open edges, the SSD connects the open ends of the ground state effectively and the state becomes {\it periodic}, having two cuts between $\Omega$ and $\bar{\Omega}$.

We have also examined EE in the other models, the antiferromagnetic $J_1$-$J_2$ chain and two-leg ladder systems under magnetic field.
The Hamiltonians are given by
\begin{eqnarray}
\mathcal{H}_{J_1{\rm-}J_2} &=& 
\sum_{j=1}^{L-1} \sum_{n=1,2} 
J_n f_{j+\frac{n}{2}} {\bm S}_{j} \cdot {\bm S}_{j+n} 
- h \sum_{j=1}^L f_j S^z_{j}, 
\label{eq:Ham-zig} \\
\mathcal{H}_{\rm lad} &=& 
J_\parallel \sum_{j=1}^{L-1} \sum_{n=1,2}
f_{j+\frac{1}{2}} {\bm S}_{n,j} \cdot {\bm S}_{n,j+1}
\nonumber \\
&&+ J_\perp \sum_{j=1}^L f_j {\bm S}_{1,j} \cdot {\bm S}_{2,j}
- h \sum_{j=1}^L f_j ( S^z_{1,j} + S^z_{2,j} ).
\label{eq:Ham-lad}
\end{eqnarray}
In Figs.\ \ref{fig:EEnt} (c)-(e), we present the DMRG results of EE for the subsystem $\Omega$ of left $l$ sites/rungs [see Fig.\ \ref{fig:EEnt} (f)].
The models in Figs.\ \ref{fig:EEnt}(c) and (e) are in critical phases with $c=1$, while the model in Fig.\ \ref{fig:EEnt}(d) has $c=2$\cite{ChitraG1997,FurusakiZ1999,HikiharaF2001,OkunishiT2003,HikiharaMFK2010}.
It is found again that the slope of EE is doubled by the SSD.
We note that, for the $J_1$-$J_2$ chain with large $J_2/J_1$, sizable boundary oscillations is observed in EE and one-point functions (not shown) even in the system with SSD.
This is presumably attributed to an effective boundary field which can not be eliminated completely by the SSD in Eq.\ (\ref{eq:Ham-zig}).\cite{correction}
However, we emphasize that even in that case the doubled slope of EE is observed, which suggests that the SSD works also for those models to lead to the topology change of the ground state.
We thus conclude that the change of slope of EE is not peculiar to a specific model but a general outcome of SSD when applied to a critical model.


\begin{figure}
\begin{center}
\includegraphics[width=65mm]{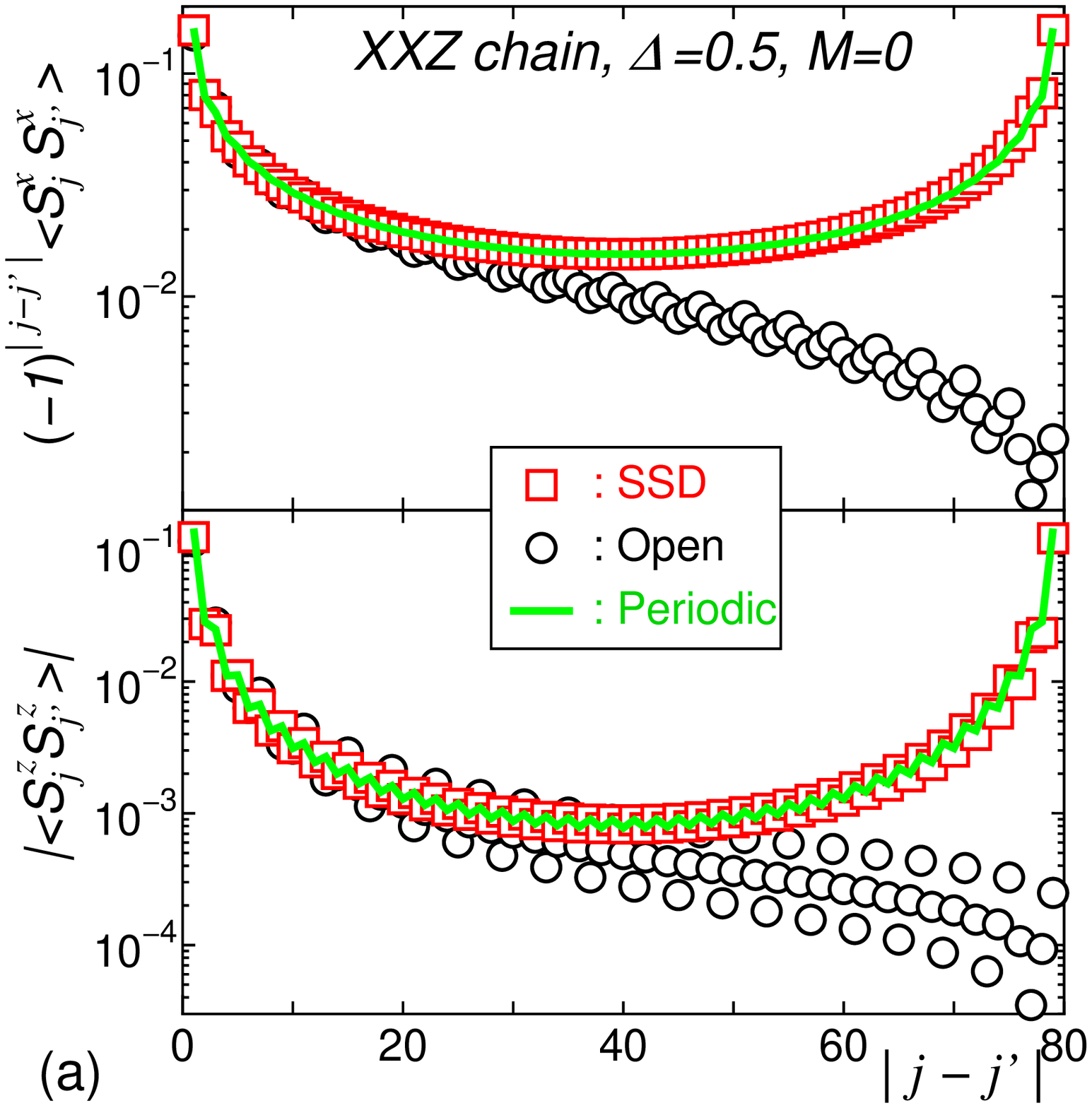}
\includegraphics[width=60mm]{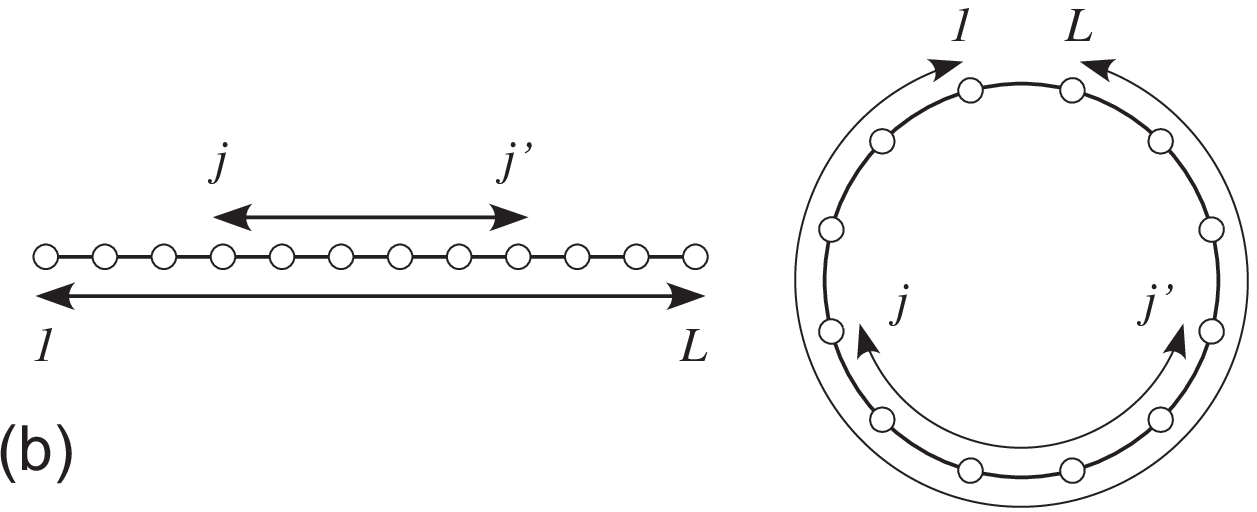}
\caption{
(a) Spin correlation functions $\langle S^\alpha_j S^\alpha_{j'} \rangle$ ($\alpha=x,z$) in the $XXZ$ chain for $L=80$ and  $(\Delta, M)=(0.5,0)$ as a function of the distance $|j-j'|$, where the sites $(j, j')$ are selected as $j = L/2 - [r/2]$ and $j' = L/2 + [(r+1)/2]$.
Squares and circles represent the DMRG data for the open chain with SSD and the uniform open chain, respectively, while the lines show the analytic result for the uniform periodic chain.
(b) Schematic picture to show the relation between the pairs $(j,j')$ in the open chain with SSD and in the periodic chain.
}
\label{fig:cor}
\end{center}
\end{figure}

\begin{figure}
\begin{center}
\includegraphics[width=65mm]{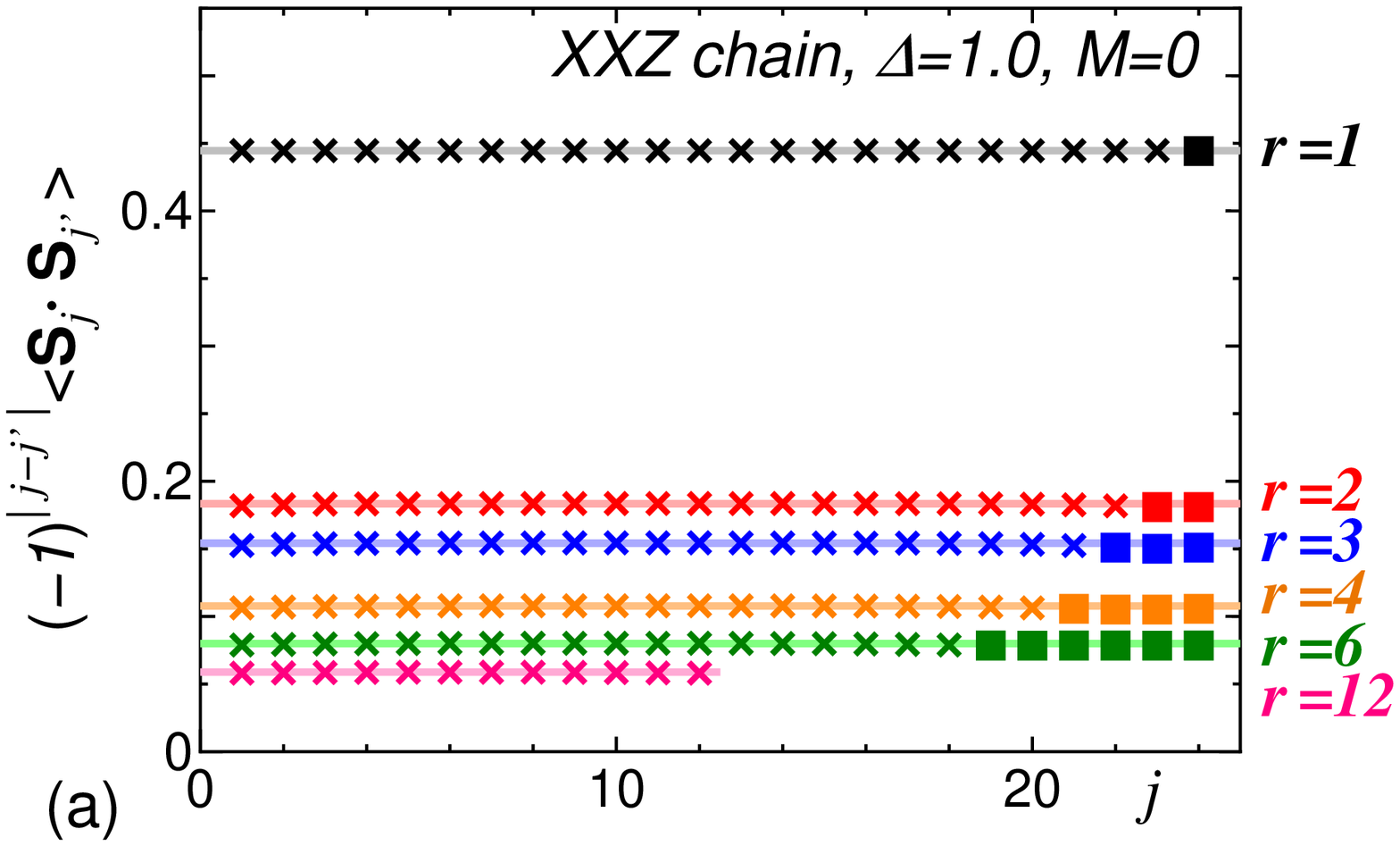}
\includegraphics[width=60mm]{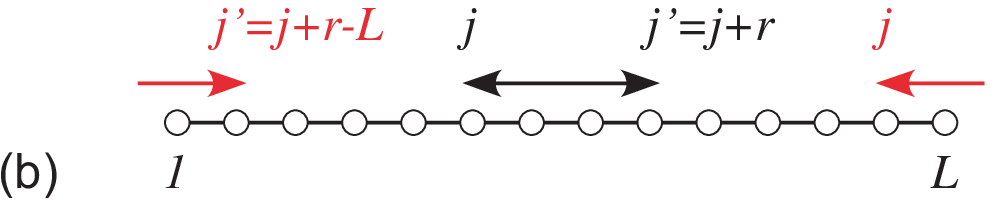}
\caption{
(a) Spin correlation function $(-1)^r \langle {\bf S}_j \cdot {\bf S}_{j'} \rangle$ with $j'=j+r$ (mod $L$) in the $XXZ$ chain for $L=24$ and  $(\Delta, M)=(1.0, 0)$ as a function of $j$ and $r$.
Symbols show the data for the open chain with SSD; 
crosses represent the correlations between the sites $j$ and $j'=j+r$ (pairs ``within" the chain), while squares are those between $j$ and $j'=j+r-L$ (pairs ``across" the edges).
Lines show the values of the correlations in the uniform periodic chain.
(b) Schematic picture to show the two sites $(j,j')$ at the ``distance" $r$.
}
\label{fig:corED}
\end{center}
\end{figure}

{\it Correlation functions: }
We next investigate the two-spin correlation functions. 
Here, we consider the spin-1/2 $XXZ$ chain in the critical regime, for which the asymptotic forms of the correlation functions are known to be
\begin{eqnarray}
\langle S^x_0 S^x_r \rangle &=& 
A^x_0 \frac{(-1)^r}{r^\eta} - A^x_1 \frac{\cos(Q r)}{r^{\eta + 1/\eta}} 
+ \cdots,
\label{eq:cxx-XXZ} \\
\langle S^z_0 S^z_r \rangle - M^2 
&=& -\frac{1}{4\pi^2 \eta r^2} + A^z_1 \frac{(-1)^r \cos(Q r)}{r^{1/\eta}} 
+ \cdots,
\label{eq:czz-szsz-XXZ}
\end{eqnarray}
where $Q=2\pi M$.
The exponent $\eta$ and the amplitudes $A^x_0$, $A^x_1$, and $A^z_1$ have been obtained as a function of $\Delta$ and $M$.\cite{Bethe-ansatz,HikiharaF2001,Lukyanov,HikiharaF}
Figure\ \ref{fig:cor} shows the DMRG results of the ground-state correlation functions in the $XXZ$ chain (\ref{eq:Ham-XXZ}) with SSD.
We also plot the DMRG data for the uniform open chain as well as the analytic result for the uniform periodic chain, the latter is obtained by replacing $r$ in Eqs.\ (\ref{eq:cxx-XXZ}) and (\ref{eq:czz-szsz-XXZ}) with $\frac{L}{\pi} \sin\left( \frac{\pi |j-j'|}{L}\right)$.
As seen in the figure, the results for the open chain with SSD agree almost completely with those for the periodic chain.

Figure\ \ref{fig:corED} (a) shows the ground-state correlation function $\langle {\bf S}_j \cdot {\bf S}_{j'} \rangle$ in a small system calculated by the exact diagonalization.
The data are plotted as a function of the position $j$ and the ``distance" $r = \min(|j-j'|, L-|j-j'|)$ [see Fig.\ \ref{fig:corED} (b)].
We observe again that the correlations in the open chain with SSD are in excellent agreement with those in the uniform periodic chain;
The results are independent of the position $j$ and, more remarkably, the correlations between the sites $j$ and $j'=j+r-L$, which locate at the distance $r$ {\it across} the open ends, have the same value as those in the periodic chain.\cite{odd-L}
We have observed the same phenomena as those in Figs. \ref{fig:cor} and \ref{fig:corED} for several parameter sets of ($\Delta, M$).
The results indicate that correlation functions, and presumably all observables, in the ground state of the systems with SSD become equal to those in the uniform periodic systems.

We note that for the two-leg ladder with zero magnetization $M=0$, which has an energy gap above the singlet ground state, the spin correlation decays exponentially even in the systems with SSD and no recovery of the correlation between edge spins is observed.
This suggests that the SSD does not work for the spin-gapped systems.

{\it Wave Functions: }
Finally, we discuss the overlap of the ground-state wave functions.
Using the exact diagonalization method, we have calculated the ground-state wave function $|{\bf v}_{\rm SSD}\rangle$ in the $XXZ$ chain (\ref{eq:Ham-XXZ}) with SSD for $L \le 24$ and several sets of $(\Delta, M)$, and compared it with the ground-state wave function $|{\bf v}_{\rm PBC}\rangle$ of the uniform periodic chain.
We have then found that the overlap of those ground-state wave functions is very close to unity;
the deviation from the unity is at most $|1 - \langle {\bf v}_{\rm SSD} | {\bf v}_{\rm PBC} \rangle | \lesssim 10^{-3}$ and exactly zero within the numerical accuracy of $10^{-14}$ for the $XX$ case ($\Delta = 0$).
The result indicates that the ground states $|{\bf v}_{\rm SSD}\rangle$ and $|{\bf v}_{\rm PBC}\rangle$ are equivalent at the level of wave function.\cite{odd-L}

We note that the equivalence of the ground-state wave functions is not trivial even in the case of the $XX$ chain [Eq.\ (\ref{eq:Ham-XXZ}) with $\Delta = 0$].
Through the Jordan-Wigner transformation, the $XX$ chain is mapped into the free fermion system and the one-particle eigenstates of the periodic chain is the simple plane waves.
On the other hand, the Hamiltonian of the open chain with SSD is not translationally invariant and its one-particle eigenstates are distinct from the plane waves.
Nevertheless, when and only when the fermions are filled up to the Fermi level, the Slater determinants of the two ground states become equivalent.
This means that the excitation spectrum and dynamics of the system with SSD are in general different from those in the periodic system.

{\it Concluding remarks: }
In summary, we have studied the SSD applied to 1D critical spin systems.
From numerical analyses of the entanglement entropy, correlation functions, and wave-function overlap, we have shown that the ground state of the open system with SSD is equivalent to the one of the uniform periodic system.

We note that our finding that the SSD realizes the periodic ground state is not restricted to a specific model but a generic feature of SSD.
We have found the change in the slope of the entanglement entropy for several spin systems and the suppression of boundary effects by SSD has also been observed in the free and interacting fermion systems\cite{GendiarKN2009,GendiarN2010}.
Only a condition required is that the system should be critical.
This may suggest that the result can be understood in a theory applicable to general critical systems, such as the conformal-field theory.
Investigating the effects of SSD on the low-energy excited states would reveal clues about the mechanism of SSD.

The results of the present study offer a new scheme to modify and further control the topology of the quantum states by the energy-scale deformation, even under the condition that the spatial shape of the system is fixed.
The approach might be applicable to real systems such as ultracold atoms, for which the spatial modulation of interatomic interactions has been demonstrated.\cite{coldatom}
Searching for an energy-scale deformation to yield the other topological change of the ground state must be another interesting problem.

\acknowledgments

We thank Shunsuke Furukawa, Akira Furusaki, Andrej Gendiar, Hosho Katsura, and Tsutomu Momoi for stimulating discussions. 
T.H. was supported in part by the MEXT and JSPS, Japan through Grand Nos.\ 21740277 and 22014016.


\begin{thebibliography}{99}
\bibitem{M-hrelation}
Unlike the usual uniform systems, the Zeeman term in the Hamiltonian with SSD does not commute with the exchange term and the relation between the field $h$ and the ground-state magnetization $M$ is not trivial.
In our calculation, we use the value of $h(M)$ obtained from the Bethe ansatz or the DMRG analysis for the corresponding uniform systems as an approximation of the field to realize the ground state with $M$ in the system with SSD.
\bibitem{PBC-SSD}
To be precise, applying SSD to a periodic chain with only nearest-neighbor interactions leads to the same Hamiltonian (\ref{eq:Ham-XXZ}) since $f_{L+\frac{1}{2}}=0$.
In this paper, we call the model (\ref{eq:Ham-XXZ}) the ``open" chain with SSD as it apparently has the open ends.
\bibitem{correction}
The reason why the SSD fails to eliminate boundary oscillations in the $J_1$-$J_2$ chain can be understood if one consider the system with $J_1=0$, which consists of two Heisenberg chains placed with a shift of a half lattice spacing.
In the case, we should apply the SSD to each chain as in Eq.\ (\ref{eq:Ham-XXZ}).
This suggests that for finite $J_2/J_1$ one must modify the form of SSD around the open edges from Eq.\ (\ref{eq:Ham-zig}).
To find the correct form is left for future studies.
\end{thebibliography}
\end{document}